\documentclass[ twocolumn,
		        aps,
		        superscriptaddress,
		        pra
		       ]{revtex4-1}
		
\usepackage[colorlinks=true,urlcolor=blue,citecolor=blue,linkcolor=red]{hyperref}
\usepackage{graphicx}
\usepackage{wrapfig}
\usepackage{booktabs}
\usepackage{amsmath}
\usepackage{amssymb}
\usepackage{braket}
\usepackage[dvipsnames]{xcolor}
\usepackage[normalem]{ulem}

\begin{document}
\title{Neural-network approach for identifying nonclassicality from click-counting data}

\author{Valentin Gebhart}\email{gebhart@lens.unifi.it}
\affiliation{QSTAR, INO-CNR, and LENS, Largo Enrico Fermi 2, I-50125 Firenze, Italy}
\affiliation{Universit\`a degli Studi di Napoli ”Federico II”, Via Cinthia 21, I-80126 Napoli, Italy}
\author{Martin Bohmann}
\affiliation{QSTAR, INO-CNR, and LENS, Largo Enrico Fermi 2, I-50125 Firenze, Italy}
\affiliation{Institute for Quantum Optics and Quantum Information - IQOQI Vienna, Austrian Academy of Sciences, Boltzmanngasse 3, 1090 Vienna, Austria}

\begin{abstract}
    Machine-learning and neural-network approaches have gained huge attention in the context of quantum science and technology in recent years.
    One of the most essential tasks for the future development of quantum technologies is the verification of nonclassical resources.
    Here, we present an artificial neural network approach for the identification of nonclassical states of light based on recorded measurement statistics.
    In particular, we implement and train a network which is capable of recognizing nonclassical states based on the click statistics recorded with multiplexed detectors.
    We use simulated data for training and testing the network, and we show that it is capable of identifying some nonclassical states even if they were not used in the training phase.
    Especially, in the case of small sample sizes, our approach can be more sensitive in identifying nonclassicality than established criteria which suggests possible applications in presorting of experimental data and online applications.
\end{abstract}

\maketitle

\section{Introduction}

    We are currently in the so-called second quantum revolution \cite{dowling_2002}, meaning that we are systematically employing quantum systems and properties in order to develop new technologies.
    For the development of quantum technologies, it is crucial to be able to characterize quantum states in order to identify quantum resources for applications.
    In this context, it is important to certify that a quantum state is nonclassical, i.e., that it cannot be described by a (semi)classical theory.
    In the field of quantum optics, nonclassicality is defined through the negativities of the Glauber--Sudarschan $P$ representation \cite{glauber_1963,sudarshan_1963}.
    In recent works \cite{streltsov_2017,yadin_2018}, it was shown that nonclassicality is, indeed, a resource for quantum technologies and can, for instance, be transformed into entanglement \cite{vogel_2014,killoran_2016}.
    In many practical applications it is therefore important to verify nonclassicality for measurement data.
    
    In this matter, an important task is the characterization of light in the few-photon regime.
    Due to the lack of photon-number-resolving detectors, so-called multiplexing strategies \cite{paul_1996, kok_2001,fitch_2003,achilles_2003,castelletto_2007,schettini_2007} have been developed which allow us to gain some insights about the measured quantum state even when a photon-number-resolving measurement is not accessible. 
    It is important to stress that such strategies do not provide direct access to the photon-number distribution, and improper interpretation of the measured statistics might lead to false certification of nonclassicality \cite{sperling_2012}.
    Therefore, it is sensible to formulate nonclassicality criteria which can be directly applied to the recorded click-counting statistics \cite{sperling_2012a,sperling_2013,luis_2015,filip_2013,lee_2016,bohmann_2019}.
    Such approaches have been successfully implemented in various experimental settings and for different quantum states of light \cite{bartley_2013,sperling_2015,heilmann_2016,sperling_2016,bohmann_2017,bohmann_2017b,kroeger_2017,bohmann_2018,obsil_2018,tiedau_2018}.

    In recent years, the methods of machine learning have been widely applied by the quantum science community \cite{dunjko_2018}. 
    As a general tool for optimization problems, machine learning was used in the study of quantum tomography \cite{torlai_2018,lee_2018,palmieri_2019,rocchetto_2019} and quantum state discrimination \cite{sentis_2015,you_2019}, quantum metrology \cite{hentschel_2010,Knott_2016,cimini_2019}, quantum error correction \cite{varsamopoulos_2017,baireuther_2018,nautrup_2019,fosel_2018}, quantum many-body systems \cite{van_nieuwenburg_2017,carleo_2017,carrasquilla_2017}, quantum state and gate preparation \cite{bisio_2010,krenn_2016,melnikov_2018,sabapathy_2019,odriscoll_2019}, and certification of quantum dynamics \cite{wiebe_2014,agresti_2019}, to mention just a few.
    Also in the task of identifying nonlocal correlations, machine learning was shown to prove advantageous \cite{deng_2018,canabarro_2019,krivachy_2019}. 
    For a recent review on machine learning and artificial intelligence in the quantum domain we refer to Ref. \cite{dunjko_2018}.
    
    In this paper, we propose and implement a machine-learning approach for the identification of nonclassical states based on their experimentally accessible click-counting statistics.
    In particular, we implement a dense artificial neural network and train the network via supervised learning with simulated click-counting data for different classical and nonclassical states.
    After the training phase, the performance of the neural network is analyzed and its results are compared to a moment-based approach for certifying nonclassicality.
    We focus special attention on the case of small sample sizes, simulating both ideal detection with unit quantum efficiency and a realistic quantum efficiency of $\eta=0.6$.
    We find that the network can identify nonclassical states even for significantly small sample sizes.
    Especially, in the case of realistic detection efficiencies, the neural-network approach performs better than the moment-based test.
    Furthermore, we show that the network can identify nonclassical states even if these states were not used in the training of the network.
    We can conclude that the neural network provides a new avenue for a fast indication of nonclassicality for small sample sizes.

    This paper is structured as follows.
    In Sec. \ref{sec:preliminaries}, we recall  some background information about click-counting detection which we will use throughout the paper.
    In Sec. \ref{sec:DNN}, we describe the design and the training of the neural network for identifying nonclassical states from their click-counting statistics.
    We apply the trained network to different simulated data sets and analyze its performance in Sec. \ref{sec:results}.
    In Sec. \ref{sec:conclusions}, we summarize and conclude the results of our paper.

\section{Preliminaries: click-counting detection}
\label{sec:preliminaries}
    
	\begin{figure}[t]
		\center
		\includegraphics[width = 0.9\columnwidth]{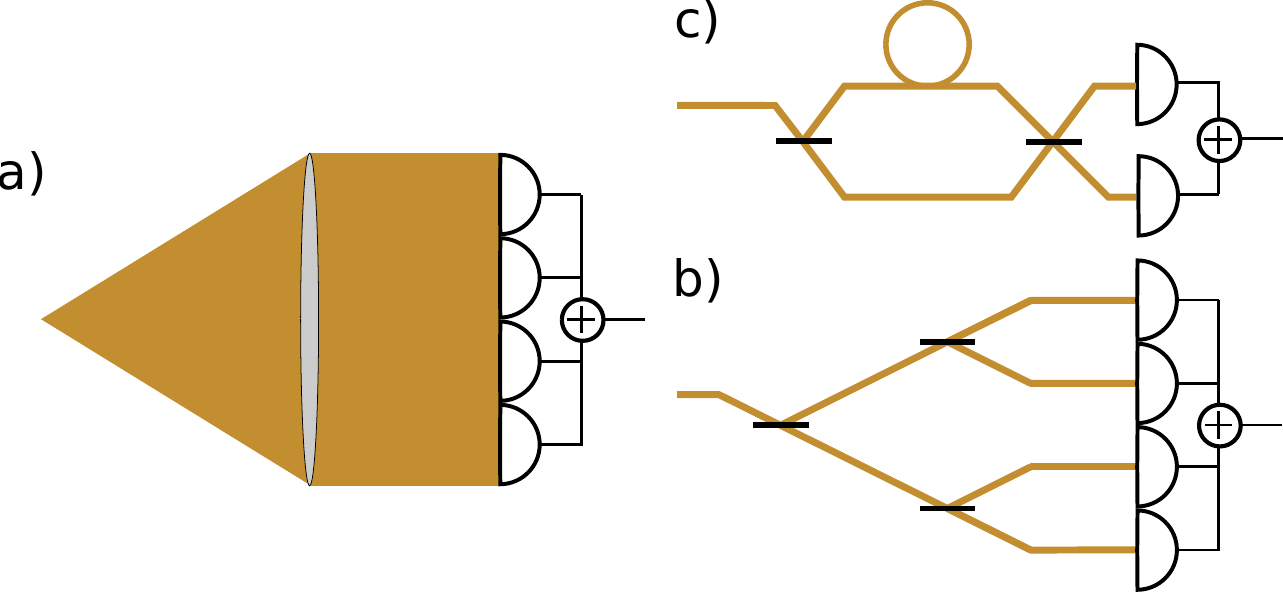} 
		\caption{
		    Schematics of different click-counting detector systems: (a) array detector, (b) spatial multiplexing, and (c) time-bin multiplexing.
		    All methods split the incoming light onto $N$ on-off detectors. 
		}
		\label{fig:multiplexing}
	\end{figure}
	
	In this section, we briefly recall the theory of click-counting detection and show how nonclassicality can be certified from the measured statistics.
	Different realizations of such click-counting detectors are shown in Fig. \ref{fig:multiplexing}.
	All different implementations have in common that they split the incoming light into $N$ detection modes of equal intensity and each mode is recorded with an on-off detector.
	The theoretical description of the click-counting statistics recorded by such devices, i.e., the probability of recording $k$ coincident clicks $(0\leq k\leq N)$, is given by the quantum counterpart of the binomial distribution \cite{sperling_2012},
	\begin{align}\label{eq:ClickStatistics}
		p_{k}=\left\langle{:}\binom{N}{k}\hat m^{N-k}\left(\hat 1-\hat m\right)^{k}{:}\right\rangle,
	\end{align}
	where  ${:}\,\cdot\,{:}$ indicates the normal-ordering prescription \cite{vogel_2006} and $\hat m$ is the no-click operator.
	Note that for typical on-off detectors, the no-click operator has the form $\hat m=\exp(-\eta \hat n/N)$ which is a function of the photon-number operator $\hat n$ and characterized by the detector's quantum efficiency $\eta$ (cf., e.g., Ref. \cite{bohmann_2017}).
	
	In what follows, we will design, implement, and test neural networks for identifying nonclassicality which take recorded click-counting statistics as inputs.
	To evaluate the performance of the neural network, we compare its predictions with a well-established method of certifying nonclassicality from click-counting data.
	In particular, we consider the approach based on the matrix of click moments \cite{sperling_2013} which has already been used in various experimental realizations \cite{bartley_2013,sperling_2015,heilmann_2016,bohmann_2017b,kroeger_2017}.
	The matrix of moments $M^{(K)}$ is defined as
	\begin{align}\label{Eq:ClMoM}
		 M^{(K)}{=}(\langle{:}\hat m^{s{+}t}{:}\rangle )_{(s,t)},
	\end{align}
	with $s,t=0,\ldots, K/2\leq N/2$ for even $K$ and $N$. 
	The superscript $(K)$ defines the highest moment within the matrix $M$ which is bounded from above by the number of detection bins $(K\leq N)$.
	Importantly, $M^{(K)}$ is non-negative for any classical light field.
	Note that the required moments and their statistical errors can be directly sampled from the click-counting statistics \cite{sperling_2013,sperling_2015}.
	The simplest form of the matrix of click moments is given by moments up to the second order,
	\begin{equation}\label{eq:secondorder}
		M^{(2)}=\begin{pmatrix} 1 & \langle{:}\hat m{:}\rangle \\ 
		\langle{:}\hat m{:}\rangle & \langle{:}\hat m^2{:}\rangle \end{pmatrix}.
	\end{equation}
	If the matrix $M^{(2)}$ is not positive semi-definite, the detected light field is nonclassical. 
	Thus, the studied state is found to be nonclassical if the minimal eigenvalue of the matrix has a significant negative value.
	We will use this condition to compare the moments method to the predictions given by the neural network.

\section{Implementation and training of the artificial neural network}
\label{sec:DNN}

    Due to the probabilistic nature of quantum physics, in order to gain significant information about quantum states, usually many repetitions of prepare and measure protocols are required. 
    In this paper, we aim at identifying nonclassical quantum states from click-counting data with small sample sizes which allows for a fast and efficient classification of quantum states. It is known that machine learning can perform well in categorizing data from small sample sizes  \cite{nielsen_2015}.
    Different kinds of machine-learning methods can be considered for the problem of detecting nonclassicality of states.
    Here, we discuss the case of supervised learning of a dense artificial neural network; that is, we train a network with data simulated from known classical and nonclassical states and test whether the network is able to recognize nonclassicality of new states which can be similar to or different from the states used in the training.  

\subsection{Implementation of the network}
    The elements of the click-counting statistics constitute the input of the neural network. 
    We will focus on the case of $N=16$ detectors.
    In this case, we have $17$ input neurons, where the $k$th neuron is given the value of the relative frequency of recording $k$ simultaneous detector clicks.
    The python libraries \textit{keras} and \textit{tensorflow} are used to implement the networkand the training. 
    
    \begin{figure}[ht]
		\center
		\includegraphics[width = 0.8\columnwidth]{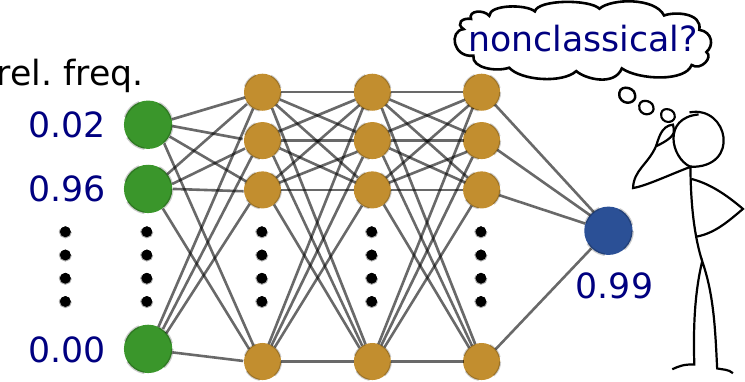}
		\caption{
		    Sketch of the dense artificial neutral network trained to identify nonclassical states of light from the multiplexed click-counting statistics. The input layer with $17$ neurons is connected to the output neuron via three fully connected layers each consisting of $50$ neurons. 
		    Nonclassicality is indicated by a high output value of the network.
		}
		\label{fig:neural_network}
	\end{figure}

\subsection{Creation of training data}

    To simulate the relative click-counting frequencies, we sample from the click-counting probabilities $p_k$ [see Eq. (\ref{eq:ClickStatistics})]. 
    As training data, we use the sampled statistics from coherent and thermal states (classical) and from Fock and squeezed states (nonclassical). 
    The exact expressions of $p_k$ for the different states are given in the Appendix. 
    Averaging over a sample size $m$ of measurement realizations generates one input data set for the neural network. 
    We generate $1000$ data points for each family of training states.
    The points are generated from states with average photon numbers $\bar n$ uniformly chosen from $\bar n \in  [1,16]$.
    The simulated data are divided into 80\% training and 20\% validation data.

\subsection{Training the network}

    We train the network for sample sizes $m=1000$ and $m=100$ and for quantum efficiencies $\eta=1$ and $\eta=0.6$. 
    Different network architectures were considered by varying the number of hidden layers, neurons per layer activation function, and optimization algorithms. 
    Optimal (minimal) architectures slightly depend on the specific quality of the training data (sample size $m$ and quantum efficiency $\eta$). 
    In the following, we choose a network with three hidden layers each consisting of $50$ neurons, as sketched in Fig. \ref{fig:neural_network}.
    We note that these parameters correspond to a rather compact network which allows for efficient and fast training and testing. 
    We use rectified linear units as neural activation functions in the network, while the output neuron is activated via the sigmoid function (cf. \cite{nielsen_2015}). 
    We train the network to minimize the mean-square error using the optimization algorithm Adam \cite{kingma_2014}.
    The network is trained until the validation error stops decreasing. 
    
\begin{figure}[t]
		\center
		\includegraphics[width = 0.99\columnwidth]{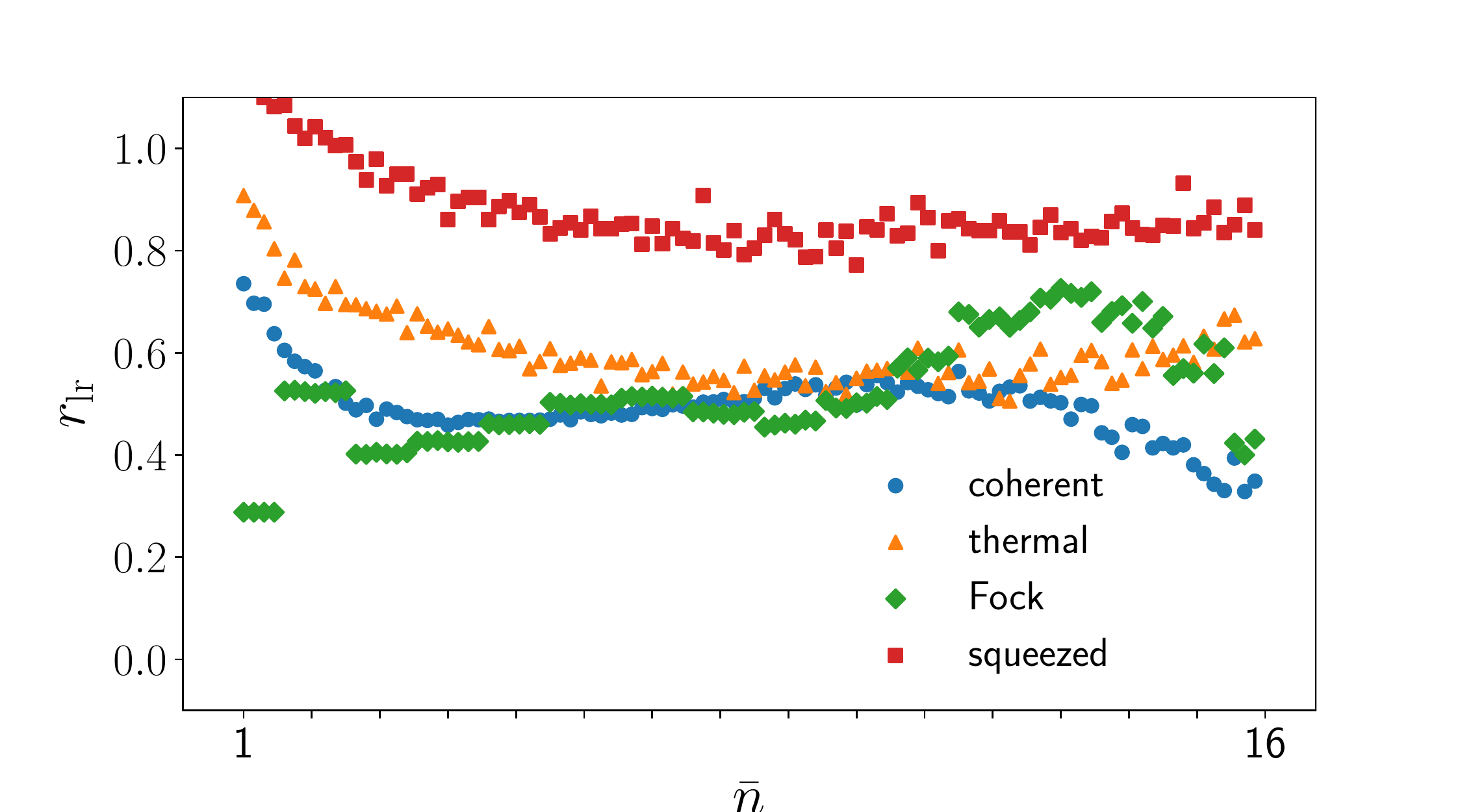}
		\caption{
		    Prediction of a linear regression of the dependence of nonclassicality on the simulated click-counting frequencies. Each color shows $100$ realizations for a different family of states, where the mean photon number $\bar n$ of the state is changed from $\bar n=1$ to $\bar n=16$. 
		}
		\label{fig:linear}
	\end{figure}

    To argue that the dependence of a nonclassicality identifier on the click-counting frequencies requires the complexity of a deep neural network, we briefly discuss the performance of linear regression as a baseline model. A linear regression of the simulated training data for $m=1000$ and ideal quantum efficiency, $\eta=1$, produces predictions as shown in Fig. \ref{fig:linear}, where for classical (nonclassical) states an output of 0 (1) was given. We observe that the prediction of the linear regression $r_\mathrm{lr}$ strongly differs from the ideal outcomes, even for the training set. 
    In particular, a linear nonclassicality identifier could never see nonclassicality of Fock states without classifying coherent and thermal states as nonclassical as well.
    Hence, from the perspective of curve fitting, it is necessary to employ a complex deep neural network in order to obtain a meaningful nonclassicality identifier.
    
\section{Certifying nonclassicality with the network}
\label{sec:results}

    The output of our neural network is a number between $0$ and $1$. 
    In our training data, we assign $0$ to all classical states and $1$ to all nonclassical states. 
    Therefore, a high output value of the network should indicate nonclassicality. 
    To quantify the performance of the network, we choose a threshold $t$ above which we say that the network identified nonclassicality.  
    The value of $t$ should be chosen such that the classical states never result in an output above $t$.
    In the following, we fix $t=0.9$. 
    While we might not identify all nonclassical states as nonclassical, this choice guarantees that it is very unlikely that a classical state is falsely identified as nonclassical, as we will see below.
    In principle, $t$ can be varied for different networks, experimental scenarios, considered classes of quantum states, or specific tasks.
    In this context, we emphasize the indicator role of the network: the output of the network will never be able to prove nonclassicality of the input.
    However, the idea is that even though we use only small sample sizes, we obtain a reliable, fast, and resource-efficient indication that an input state is nonclassical.
    Such a fast identification can be of practical importance in many scenarios, as we shall discuss in Sec. \ref{sec:conclusions}.
    
    We compare the output of the network with the nonclassicality test based on the moments of the click-counting distribution (see Sec. \ref{sec:preliminaries}). 
    This provides a comparison with established nonclassicality tests for such measurement scenarios.
    The moments method yields the minimal eigenvalue $\bar x_\mathrm{mom}$ of the matrix of moments in Eq. \eqref{Eq:ClMoM} together with its sampling error $\Delta_{x_\mathrm{mom}}$.
    If we find that the estimated minimal eigenvalue $x_\mathrm{mom}=\bar x_\mathrm{mom} \pm\Delta_{x_\mathrm{mom}}$ is significantly negative, nonclassicality is verified through the matrix of moments approach.
    A commonly accepted minimal significance level is three standard deviations.
    In our case, this means that the moments method detects nonclassicality if its value is three standard deviations below zero, i.e., $r_\mathrm{mom}:=-\bar x_\mathrm{mom}/\Delta_{x_\mathrm{mom}}>3$. 
    Such a significance level is especially important as we do not want to falsely identify classical states as nonclassical ones.
    We will use the significance value $r_\mathrm{mom}$ as a comparison with our results obtained from the neural-network approach.

\subsection{Perfect detectors} 
	
    \begin{figure*}[t]
        \includegraphics[width=.9\textwidth]{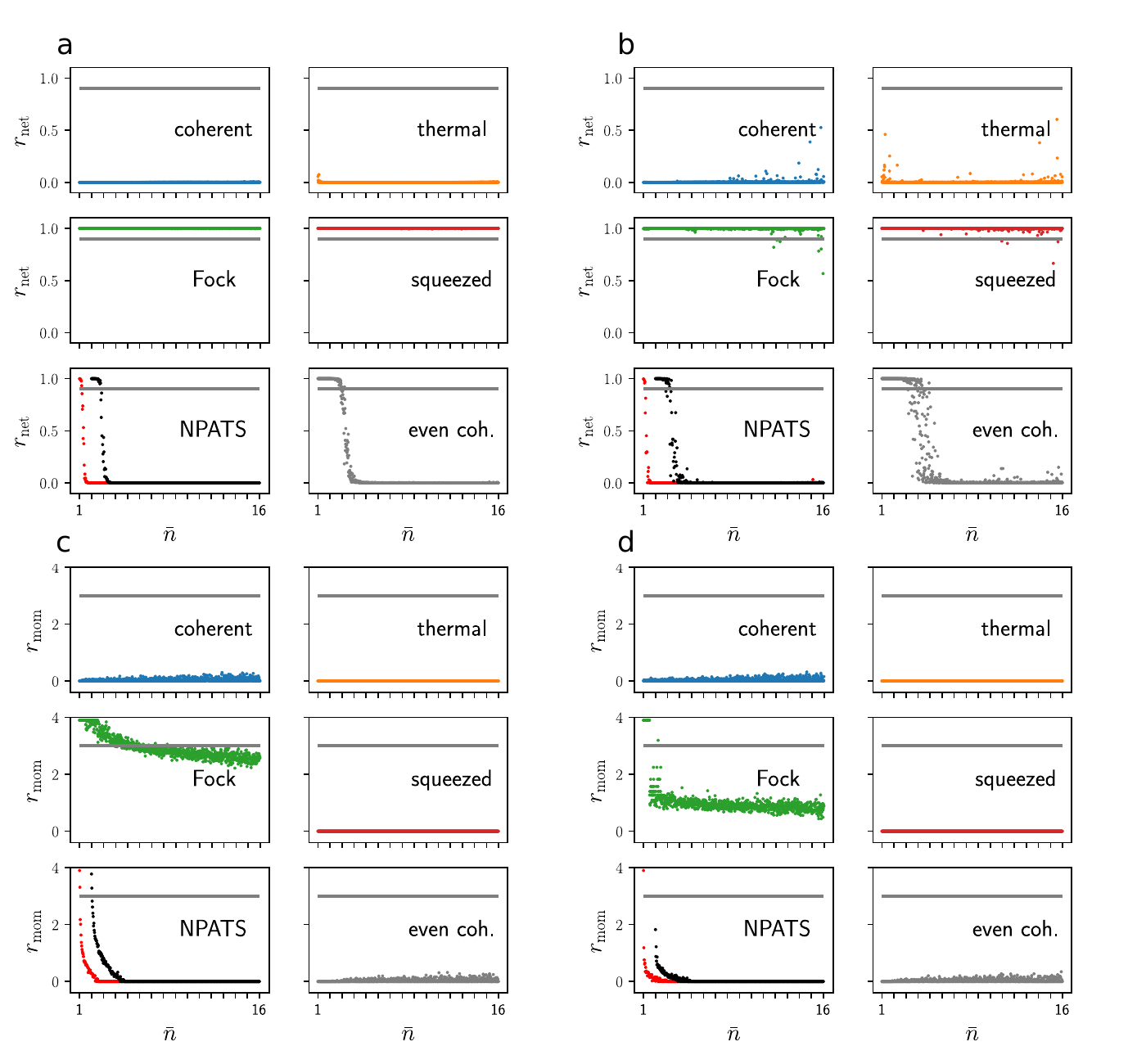}
        \caption{
        Comparison of the output of the network (a) and (b) with the nonclassicality test using the moments of the measurement distribution (c) and (d) for unit quantum efficiency of the detectors, $\eta=1$. (a) and (c) correspond to sample size $m=1000$, while (b) and (d) correspond to $m=100$. Each subplot shows $1000$ realizations for a different family of states, where the mean photon number $\bar n$ of the state is changed from $\bar n=1$ to $\bar n=16$. For the Fock states, the points for noninteger $\bar n$ correspond to the largest integer $n<\bar n$. For the NPATS, we show single-photon-added (red) and two-photon-added (black) thermal states. The gray lines represent the nonclassicality thresholds. }
        \label{fig:real_perfect_efficiency}
    \end{figure*}

    At first, we consider data obtained from detectors with an ideal unit quantum efficiency, $\eta=1$. 
    Figure \ref{fig:real_perfect_efficiency} shows the output of the network for the different families of states.
    For each family, we generate 1000 different realizations of sample size $m$ with mean photon numbers ranging from $1$ to $16$. The data are sampled from the click-counting distributions of the different states as explained in Sec. \ref{sec:DNN}.
    In Fig. \ref{fig:real_perfect_efficiency} (a), we show the performance of the network for a sample size $m=1000$, while Fig. \ref{fig:real_perfect_efficiency} (b) displays the network's output for $m=100$. 
    In each case, the nonclassicality threshold $t=0.9$ is indicated with a straight line. 
    The corresponding results of the moments method are shown in Figs. \ref{fig:real_perfect_efficiency}(c) and \ref{fig:real_perfect_efficiency}(d) together with the significance-threshold level $r_\mathrm{mom}=3$.
    (Note that, for better visualization, we set all classical values $r_\mathrm{mom}<0$ to zero.)

    In both cases ($m=1000$ and $m=100$), the network clearly recognized Fock and squeezed states as nonclassical, while coherent and thermal states are never seen as nonclassical. 
    Hence, the neural network is capable of correctly identifying the cases which were used for the training.
    For $m=100$, we observe that the results of the neural network are less clear due to the very limited sample size.
    However, it is still surprising that for such small sample sizes the network still shows a good performance.
    Lowering the sample size even further eventually results in a finite probability of falsely categorizing classical states as nonclassical, ending the network's validity as nonclassicality indicator. 
    
    Let us compare the neural network's results with the ones for the moments method.
    We see that for the two classes of classical states (coherent and thermal states), both techniques yield compatible results.
    On the other hand, there is a significant difference between the results for the cases of the nonclassical Fock and squeezed states.
    For $m=1000$, Fock states are recognized only by the moments method with an acceptable significance up to a photon number of about $8$ [see Fig. \ref{fig:real_perfect_efficiency}(c)]. 
    Furthermore, for $m=100$, nonclassicality can be seen only for Fock states up to a maximum of two photons [see Fig. \ref{fig:real_perfect_efficiency}(d)].
    We also observe that the moments-based approach is never capable of identifying squeezed states as nonclassical independent of the sample size $m$.
    This can be explained by the fact that the moments method is insensitive to the nonclassical feature of squeezed states.
    Therefore, we can conclude that the neural network can be more sensitive towards nonclassical states than the established moments method and that it may also identify nonclassical states where the latter fails completely.

    To examine how the network recognizes nonclassicality of states that have not been used in the training, different nonclassical states or mixtures of (nonclassical) states (see also Sec. \ref{sec:mixtures}) can be tested. 
    Here, we test the network with $n$-photon-added thermal states (NPATS) and with even coherent states (see Fig. \ref{fig:real_perfect_efficiency}).
    For the NPATS, only states which are close to one- and two-photon added vacuum states are recognized.
    This can be understood by recognizing that NPATS are very similar to thermal states for an increasing mean photon number.
    Hence, the network has difficulties distinguishing NPATS from thermal states for larger mean photon numbers. 
    Similarly, even coherent states are recognized as nonclassical only for mean photon numbers $\bar n \lesssim 3$. 
    For larger mean photon numbers, the two coherent states in the superposition $(\left| \alpha \right\rangle+\left| -\alpha \right\rangle)$ become nearly orthogonal with the consequence that the state is very similar to a mixture of two coherent states for which the network cannot recognize nonclassicality.
    
    We also trained neural networks including these additional states in the training data.
    Even then, their similarities to thermal states or mixtures of coherent states render them impossible to be fully recognizable by the network.
    Additionally, we note that the network trained with a larger sample size is able to classify test data from smaller sample sizes and vice versa.

\subsection{Realistic detection efficiencies} 

    \begin{figure*}[t]
        \includegraphics[width=.9\textwidth]{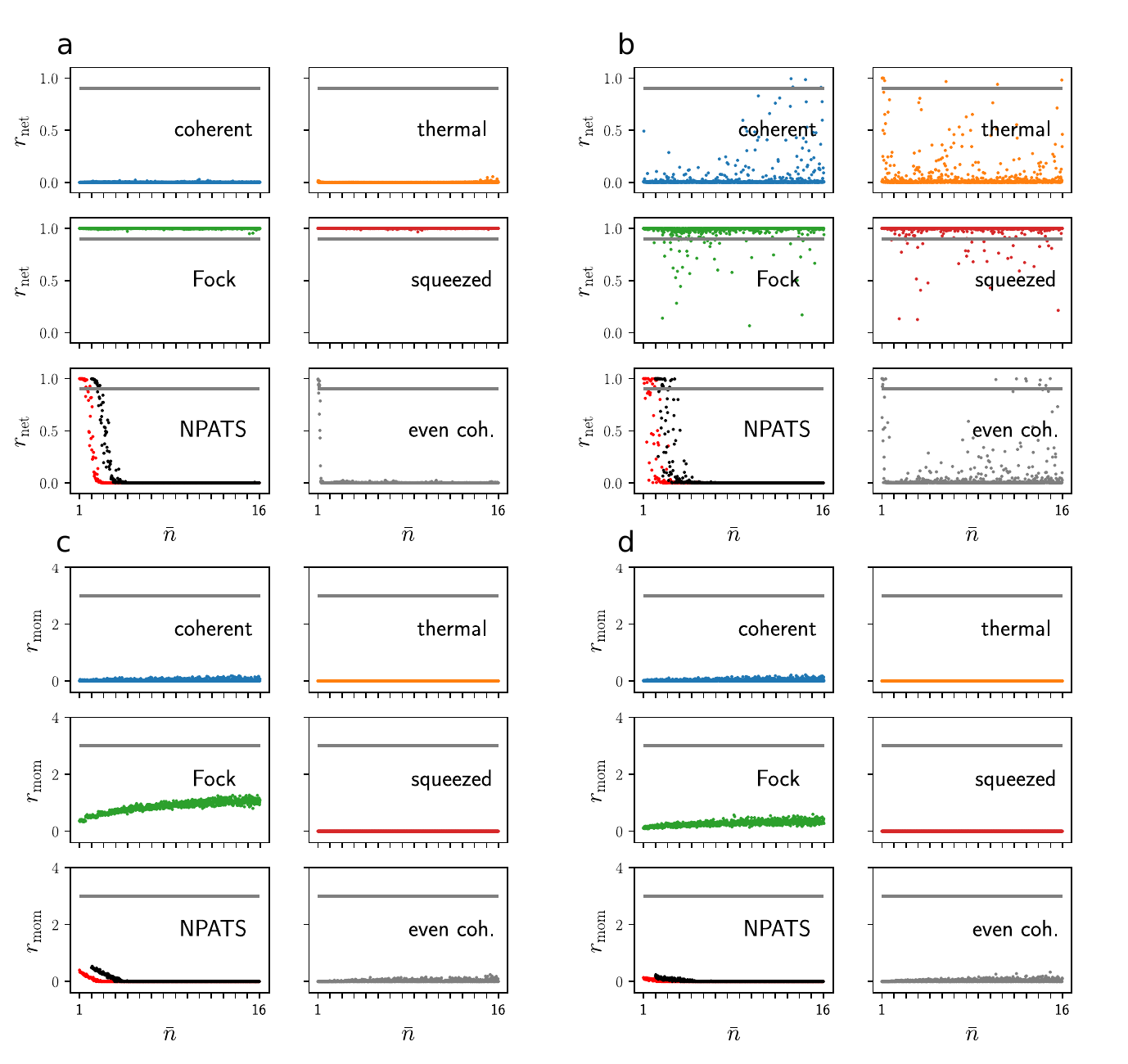}
        \caption{
        Comparison of (a) and (b) the output of the network with (c) and (d) the nonclassicality test using the moments of the measurement distribution for a realistic quantum efficiency of the detectors, $\eta=0.6$. (a) and (c) correspond to sample size $m=1000$, while (b) and (d) correspond to $m=100$. Each subplot shows $1000$ realizations for a different family of states, where the mean photon number $\bar n$ of the generating state is uniformly changed from $\bar n=1$ to $\bar n=16$. The gray lines represent the nonclassicality thresholds.}
        \label{fig:ratesandreal06}
    \end{figure*}
    
    To simulate a more realistic experiment we now consider detectors with detection efficiencies $\eta=0.6$. 
    For this purpose, we implement a neural network by training with simulated data with the same quantum efficiency.
    In Fig. \ref{fig:ratesandreal06}, in analogy to Fig. \ref{fig:real_perfect_efficiency}, we show the outputs of the network [Figs. \ref{fig:ratesandreal06}(a) and \ref{fig:ratesandreal06}(b)] and the moments methods [Figs. \ref{fig:ratesandreal06}(c) and \ref{fig:ratesandreal06}(d)] for the different families of states.
    Again, for every state we simulate 1000 different realizations with mean photon number ranging from $1$ to $16$. 
    
    We observe that for $m=1000$ [Fig. \ref{fig:ratesandreal06}(a)], the trained states are still recognized very well with slightly increased fluctuations compared to the ideal detector network [see Fig. \ref{fig:real_perfect_efficiency}(a)]. 
    Also, the NPATS are recognized in the same regime as before.
    Even coherent states, however, are recognized as nonclassical only for mean photon numbers around $\bar n=1$.
    In this case, the loss introduced by the finite detection efficiency reduces the parameter range for which the network can identify quantum correlations.
    Overall, we can conclude that the neural network approach still yields positive results for realistic quantum efficiencies.
    In comparison, the moments method [Fig.\ref{fig:ratesandreal06}(c)] yields in this case no significant certification of the nonclassical quantum states, which is due to the combination of the relatively low sample sizes and the finite quantum efficiency.
    Hence, we can conclude that the neural-network approach performs better under these conditions.
    
    The results of the neural network for very small sample sizes $m=100$ and quantum efficiency $\eta=0.6$ are shown in Fig. \ref{fig:ratesandreal06}(b).
    We see that the results are noisier in comparison to the previous example; however, the overall discrimination between classical and nonclassical states can still be observed.
    This an interesting result as it shows that the network can still be of use even for realistic quantum efficiencies and very low sample sizes.
    In comparison, the moments method yields no significant results for any of the tested nonclassical states [see Fig. \ref{fig:ratesandreal06}(d)].
    We can conclude that the neural network is advantageous under the considered conditions and that this approach can, indeed, be used for fast (pre)identification tasks with a limited amount of data.

    We note, however, that the fluctuations in the results of the neural network lead to the fact that in some rare cases classical states are falsely identified as nonclassical ones [see Fig. \ref{fig:ratesandreal06}(b); coherent and thermal states].
    Increasing the identification threshold for nonclassical states would not resolve this problem. 
    Therefore, the considered case is at the limit of the operational range of the neural-network approach.
    Decreasing the sample size or the detection efficiency further eventually terminates the validity of the nonclassicality predictions of the network.

    Note that the data stemming from detectors with a limited quantum efficiency cannot be classified by networks trained with data for perfect detector statistics. 
    This is, however, not a fundamental problem of the approach, as in practical experimental realizations the expected quantum efficiencies are usually known and it is possible to include this knowledge in the training of the network.
    A possible extension of the presented approach could be to train the network using data generated for different values of detection efficiencies, and see whether the network is able to learn to classify these different cases in parallel.  

\subsection{Testing nontrained mixed states}\label{sec:mixtures}

    Having trained the network to detect nonclassicality, different untrained states can be tested. 
    In this context, it is particularly interesting to investigate parametrized mixtures of previously trained classical and nonclassical states.
    The purpose of such an analysis is twofold.
    First, it provides the possibility of analyzing the performance of the network for an additional family of mixed states (besides the already considered cases of NPATS and even coherent states).
    Second, it allows us to study the behavior of the network when it has to classify a state which is a mixture of classical and nonclassical ones where both were used in the training.
    In particular, we can investigate whether the network performs smoothly at the transition from a purely nonclassical to a purely classical state.

    Here, as an example, we test a mixture of a coherent state $\left| \alpha \right\rangle$ and a Fock state $\left| n \right\rangle$,  
    \begin{equation}
        \rho = p  \left| \alpha \right\rangle \left\langle \alpha \right| + (1-p)  \left| n \right\rangle \left\langle n \right|,
    \end{equation}
    with $0\leq p \leq 1$ and an equal mean photon number $n=\left|\alpha\right|^2$.
    In this case, we use data and a network generated from perfect detectors, $\eta=1$, for a sample size of $m=1000$. 
    
\begin{figure}[h]
		\center
		\includegraphics[width = 0.99\columnwidth]{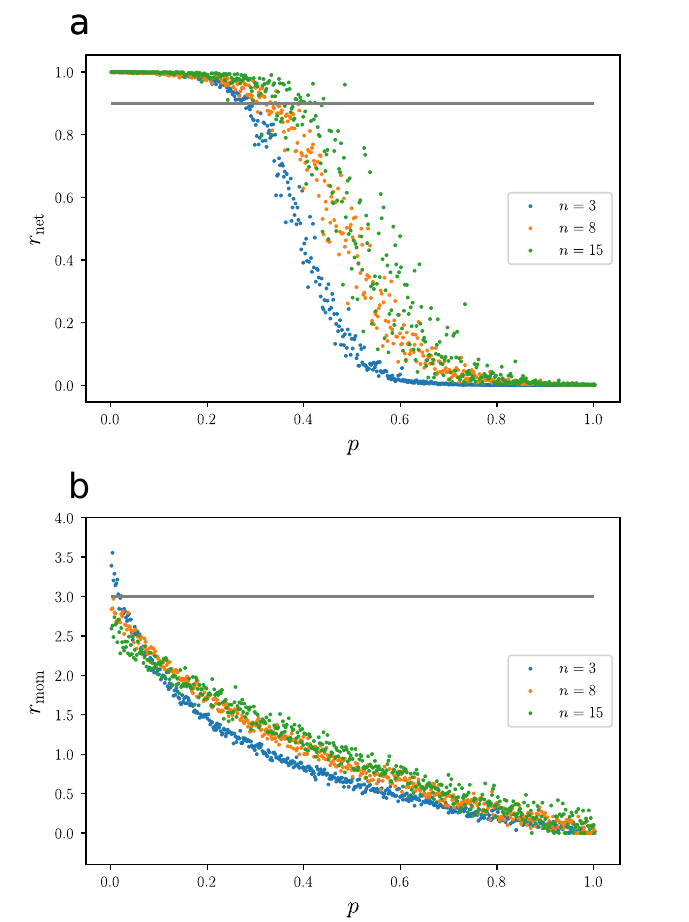}
        \caption{(a) Output of the network for the non-trained mixed state $\rho = p  \left| \alpha \right\rangle \left\langle \alpha \right| + (1-p)  \left| n \right\rangle \left\langle n \right|$ with $0\leq p \leq 1$ and $n=\left|\alpha\right|^2=3$ (blue), $n=8$ (yellow), and $n=15$ (green). (b) Corresponding significance of the nonclassicality test of the moments method. The gray lines correspond to the nonclassicality thresholds.}
		\label{fig:mixedcohnum}
	\end{figure}

    In Fig. \ref{fig:mixedcohnum}(a), we show the output of the network for different realizations corresponding to $\rho$, with $n=3,8,15$ and $0\leq p \leq 1$. 
    For small $p$, the state resembles a Fock state and is classified as nonclassical, while for large $p$, the contribution of the coherent state dominates and the state is not seen as nonclassical anymore. 
    Importantly, we observe that the transition between the nonclassical ($p=0$) and classical ($p=1$) regimes is smooth, which shows that the network can resolve this transition without showing any bias which might arise from the training.
    The network can identify nonclassicality of the mixed state up to values $p\approx 0.4$.
    We also see that in the intermediate regime, states with higher mean photon numbers $n$ can be identified as nonclassical in a wider parameter regime, which is due to the considered number of detection bins $N=16$.
    
    Additionally, we can compare the network's performance in this transition with the moments method [see Fig. \ref{fig:mixedcohnum}(b)].
    We observe that the moments method yields a significant certification of nonclassicality only for the nearly pure nonclassical case ($p\approx 0$) and that value $r_{\mathrm{mom}}$ quickly decreases with increasing $p$.
    Hence, by comparing Figs. \ref{fig:mixedcohnum}(a) and \ref{fig:mixedcohnum}(b), we see that the neural network can identify nonclassicality of the studied mixed state in a larger parameter region, which further underlines the sensitivity of the approach.
    
\section{Conclusions}
\label{sec:conclusions}
    
    We implemented and trained a dense artificial neural network for identifying nonclassical states of light based on the recorded click-counting statistics obtained from multiplexed detection systems.
    For the training and testing of the network, we simulated corresponding click-counting data for different families of classical and nonclassical quantum states of light.
    The results of the neural network were compared with nonclassicality conditions based on the moments of the recorded statistics.
    Our results show that the trained network is capable of distinguishing nonclassical states from classical ones for various types of quantum states and for different parameter regions, such as different mean photon numbers.
    In the case of small sample sizes, our machine-learning approach may be more sensitive to nonclassicality than the moment-based method.
    Importantly, the network is even capable of revealing nonclassicality of some states which were not used in the training phase. 
      
    It is important to stress that the results of the network are always just an indicator for nonclassicality.
    For a  significant and reliable certification of nonclassicality, one additionally needs to implement a nonclassicality test or witness which provides a robust verification taking into account error bars.
    However, there are various practical scenarios for which the neural network can be advantageous.
    It can be used for fast and efficient presorting of experimental data without the need for implementing and performing an elaborate data analysis.
    Here, we emphasize that the neural-network approach shows a good performance even for very small sample sizes.
    This can, for instance, facilitate the search for promising experimental parameter regions for which nonclassical states are generated. 
    Additionally, it is possible to use the neural network as an online tool during the data acquisition.
    In this way, one can check for nonclassical features directly in real time.
    This also opens the possibility to identify possible issues in the experimental implementation while performing the measurement.
    Such an online usage is especially useful whenever the data acquisition rate is rather low, such as in the case of some heralded or conditional state generation schemes.
    
    Furthermore, it is possible to optimize the training of the neural network according to each particular experimental platform.
    Usually, it is possible to predict the classes of quantum states which are generated in a certain experiment and, therefore, it is possible to adapt and optimize the training of the neural network to these conditions.
    In the same way, the detection parameters, such as the number of detection bins or the quantum efficiency, can be adapted for each specific case.
    
    Finally, we would like to point out that the presented approach could be extended in different ways.
    First, it is possible to extend the neural-network method to two-mode and multimode scenarios where each mode is recorded with a multiplexed detector system.
    This would allow us to test for cross correlations between the modes.
    Second, it is, in principle, possible to adapt the presented approach to other measurement strategies and other physical systems.
    For example, one could implement a network which is capable of identifying nonclassicality based on quadrature measurement data from balanced homodyne detectors, or one could think of adopting the presented technique to measurements in cold-atom experiments.

\section*{Acknowledgment}
    MB acknowledges financial support from the Leopoldina Fellowship Programme of the German National Academy of Science (LPDS 2019-01).

\appendix*

\begin{table*}[b]
    \centering
    \begin{tabular}{cccc}
    \hline \hline 
      State & Parameters & Probability $p_k$ & Average number $\bar n$  \\ \hline 
      coherent   & $\alpha\in \mathbb{C}$ & $\binom{N}{k}\left(e^{-\eta |\alpha|^2/N}\right)^{N-k}\left(1-e^{-\eta |\alpha|^2/N}\right)^{k}$ \cite{sperling_2012} & $|\alpha|^2$ \\   
      thermal   & $n_\mathrm{th}\in \mathbb{R}^+$  & $\frac{1}{n_\mathrm{th} +1}\sum_{j=0}^\infty \left(\frac{n_\mathrm{th}}{n_\mathrm{th} + 1}\right)^j \mathcal{D}^\eta_{k,j}$ \cite{kuehn_2014} & $n_\mathrm{th}$ \\  
      Fock   & $n\in\mathbb{N}$ & $\mathcal{D}^\eta_{k,n}$ \cite{sperling_2014}, see Eq. \eqref{eq:Dsymbol} & $n$ \\  
      squeezed   & $\xi\in\mathbb{C}$ & $\frac{1}{\cosh |\xi|}\sum_{n=0}^\infty \left(\frac{\tanh |\xi|}{2}\right)^{2n}\frac{(2n)!}{(n!)^2}\mathcal{D}^\eta_{k,2n} $ \cite{sperling_2012} & $\sinh^2 |\xi| $ \\  
      NPATS   & $n_\mathrm{th}\in\mathbb{R}^+,n\in\mathbb{N} $ & $\frac{1}{(n_\mathrm{th} +1)n_\mathrm{th} ^n}\sum_{j=n}^\infty \binom{j}{n}\left(\frac{n_\mathrm{th}}{n_\mathrm{th} + 1}\right)^j \mathcal{D}^\eta_{k,j}$ \cite{kuehn_2014}  & $n_\mathrm{th}(n+1)+n$ \cite{agarwal_1992} \\  
      even coherent   & $\alpha\in \mathbb{C}$ & $\binom{N}{k} \sum_{j=0}^{k} \binom{k}{j}(-1)^j g(N-k+j,\alpha)$ & $|\alpha|^2\frac{1-e^{-2|\alpha|^2}}{1+e^{-2|\alpha|^2}}$ \\
      \hline \hline 
    \end{tabular}
    \caption{Probability $p_k$ of $k$ detector clicks and mean photon numbers $\bar n$ for the different states, depending on the state-specific parameters. $N$ is the number of on-off detectors in the multiplexing device, and $\eta$ is the quantum efficiency of each detector.}
    \label{tab:statestatistics}
\end{table*}

\section{Click-counting statistics}

    The simulation of the click-counting data for each considered state and sample size is performed by sampling from the click-counting statistics $p_k$ in Eq. \eqref{eq:ClickStatistics}.
    In Table \ref{tab:statestatistics}, we list the explicit probabilities and mean photon numbers depending on the parameters for the different states. 
    The symbol $\mathcal{D}^\eta_{k,n}$ is defined as \cite{sperling_2014}
    \begin{equation}
        \mathcal{D}^\eta_{k,n}=\binom{N}{k}\sum_{j=0}^k \binom{k}{j}(-1)^{k-j}\left(1-\eta \frac{N-j}{N}\right)^n.\label{eq:Dsymbol}
    \end{equation}
    For the NPATS, $n_\mathrm{th}$ is the mean photon number of the initial thermal states to which $n$ photons are added.
    The probabilities $p_k$ for the even coherent states $\left|\alpha_+\right\rangle=\mathcal{N}(\left|\alpha\right\rangle+\left|-\alpha\right\rangle)$ follow directly from Eq. \eqref{eq:ClickStatistics}, where $g$ is given by \cite{lipfert_2015} 
    	\begin{align}
    		g(\lambda,\alpha):=& \left\langle\alpha_+ | {:}e^{-\lambda \hat n/N}{:} | \alpha_+\right\rangle \\ =& \frac{e^{-\lambda |\alpha|^2/N}+e^{(\lambda/N-2) |\alpha|^2}}{1+e^{-2 |\alpha|^2}}.
    	\end{align}
    The probabilities for thermal, squeezed, and photon-added thermal states include infinite sums. 
    For the simulation of the data, these sums were truncated at a sufficiently large cutoff.


\bibliography{biblio}
\end{document}